\documentclass[reprint,twocolumn,aps,superscriptaddress,amsmath,amssymb,showkeys]{revtex4-1}

\usepackage{graphicx}
\usepackage{dcolumn}
\usepackage{bm}
\usepackage{gensymb}
\usepackage[utf8]{inputenc}
\usepackage[T1]{fontenc}
\usepackage{mathptmx}
\usepackage{upgreek}

\begin{document}

\title{Silica-Silicon Composites for Near-Infrared Reflection: A Comprehensive Computational and Experimental Study}

\author{Kevin Conley}
\affiliation{Department of Applied Physics, QTF Centre of Excellence, Aalto University School of Science, P.O. Box 11100, FI-00076 Aalto, Finland}
\author{Shima Moosakhani}
\affiliation{Department of Chemistry and Materials Science, Aalto University School of Science, P.O. Box 11100, FI-00076 Aalto, Finland}
\author{Vaibhav Thakore}
\affiliation{Department of Applied Mathematics, The University of Western Ontario, London, Ontario N6A 5B7, Canada}
\affiliation{Center for Advanced Materials and Biomaterials Research, The University of Western Ontario, London, Ontario N6A 3K7, Canada}
\author{Yanling Ge}
\affiliation{Department of Chemistry and Materials Science, Aalto University School of Science, P.O. Box 11100, FI-00076 Aalto, Finland}
\author{Joonas Lehtonen}
\affiliation{Department of Chemistry and Materials Science, Aalto University School of Science, P.O. Box 11100, FI-00076 Aalto, Finland}
\author{Mikko Karttunen}
\affiliation{Department of Applied Mathematics, The University of Western Ontario, London, Ontario N6A 5B7, Canada}
\affiliation{Center for Advanced Materials and Biomaterials Research, The University of Western Ontario, London, Ontario N6A 3K7, Canada}
\affiliation{Department of Chemistry, The University of Western Ontario, London, Ontario N6A 5B7, Canada}
\author{Simo-Pekka Hannula}
\affiliation{Department of Chemistry and Materials Science, Aalto University School of Science, P.O. Box 11100, FI-00076 Aalto, Finland}
\author{Tapio Ala-Nissila}
\email{tapio.ala-nissila@aalto.fi.}
\affiliation{Department of Applied Physics, QTF Centre of Excellence, Aalto University School of Science, P.O. Box 11100, FI-00076 Aalto, Finland}
\affiliation{Interdisciplinary Centre for Mathematical Modelling, Department of Mathematical Sciences, Loughborough University, Loughborough, Leicestershire LE11 3TU, United Kingdom}

\date{\today}

\begin{abstract}
Compact layers containing embedded semiconductor particles consolidated using pulsed electric current sintering exhibit intense, broadband near-infrared reflectance. The composites consolidated from nano- or micro-silica powder have a different porous microstructure which causes scattering at the air-matrix interface and larger reflectance primarily in the visible region. The 3~mm thick composite compacts reflect up to 72\% of the incident radiation in the near-infrared region with a semiconductor microinclusion volume fraction of 1\% which closely matches predictions from multiscale Monte Carlo modeling and Kubelka-Munk theory. Further, the calculated spectra predict an improvement of the reflectance by decreasing the average particle size or broadening the standard deviation. The high reflectance is achieved with minimal dissipative losses and facile manufacturing, and the composites described herein are well-suited to control the radiative transfer of heat in devices at high temperature and under harsh conditions. 
\end{abstract}


\maketitle

\section{Introduction}\label{intro}
Controlled propagation of near-infrared (NIR) electromagnetic waves has been demonstrated in diverse applications, including biosensing~\cite{kabashin2009plasmonic}, thermal energy management~\cite{jonsson2017solar}, and switchable meta-mirrors~\cite{ma2018tunable}. The NIR region is the principal component of thermal radiation and accounts for 52\% of the sun's irradiance power~\cite{sousa2017hybrid}. Developing compact layers to trap light in the NIR could be used to provide quantitative spectral information about heat transport under harsh conditions~\cite{mityakov2012gradient} or increase efficiency and reduce cost in solar cells and thermal energy management~\cite{siebentritt2017chalcopyrite,yin2015integration,van2015light}.

Materials or devices employed in thermal insulation and Gradient Heat Flux sensing applications are subjected to high temperature and harsh conditions. Metallic particles are typically used in many applications because of their strong plasmonic response, but they suffer from intrinsic Ohmic losses~\cite{baffou2013thermo}. In contrast, semiconductors have low dissipative losses and the quality of the scattering resonances is better maintained than in metallic particles at high temperatures~\cite{thakore2019thermoplasmonic}.  

Composite compacts embedded with low volume fraction of low bandgap semiconductor microparticles, such as silicon or germanium, are predicted to reflect over 80\% of NIR solar radiation~\cite{conley2020directing} and up to 65\% of blackbody radiation~\cite{tang2017plasmonically}. The strong reflectance is achieved with much lower volume fraction (1\%) than compacts of similar thickness containing hBN-platelets at 50\% volume fraction~\cite{slovick2015tailoring}. The light interacts with the particles to generate strong scattering resonances. Unique optical effects such as unidirectional scattering and enhanced Raman scattering arise when there is interference between the resonances and strong localization of the electric and magnetic fields~\cite{kerker2013scattering,alaee2015generalized,kruk2017functional}. 

Consolidating microparticles into a matrix (host) can be achieved by various techniques such as sol–gel~\cite{maduraiveeran2007gold}, vapor phase axial deposition~\cite{cuevas2000effect}, pressureless sintering of floc-casting~\cite{hiratsuka2007fabrication}, slip-casting~\cite{szafran2007ceramic}, and electric current assisted sintering (Pulsed electric current sintering, PECS or spark plasma sintering, SPS)~\cite{munir2011electric}. Among these wide ranging techniques, the rapid processing time of PECS is very effective at limiting diffusive transformations such as crystallization, oxidation, and grain growth even at elevated processing temperatures. On the other hand, this technique provides a facile route to achieve the required thickness in comparison with other methods like sol-gel that would need successive layer-by-layer deposition. Typical heating rates in PECS range from 5\degree C~/~min up to 2000\degree C~/~min, and the typical sintering times are in minutes~\cite{munir2011electric}. PECS has been used in the consolidation of SiO$_2$ composite layers. Singh et al. synthesized a silica/graphene oxide (GO) composite with a GO content ranging from 0.001 to 10\%~\cite{singh2014tuning}. The silica/GO compacts were consolidated at a temperature of 1200\degree C at a heating rate of 100\degree C~/~min with a hold time of 1 minute at a pressure of 50 MPa. He et al. made a SiO$_2$/SiC composite using SiC deposited by use of PVD onto SiO$_2$ particles~\cite{he2013synthesis}. They observed that the densification during the PECS began at 1127\degree C of SiO$_2$ with the highest density obtained at 1277\degree C and above, but the NIR reflectivity of the compact layers prepared by PECS was not examined.

In this study, highly dense SiO$_2$ compacts embedded with silicon particles are synthesized by PECS and their diffuse reflectance spectra are compared with results from Monte Carlo simulations and Kubelka-Munk theory from the mid-ultraviolet to NIR. The synthesis, characterization, and computational methodologies are described in Section~\ref{methods}. The porosity and microstructure of the experimentally prepared silica-silicon particle composites are described in Section~\ref{exp}. In Section~\ref{spectra}, the optical properties of silica-silicon composite compacts of various thickness are calculated and compared with experiments. The optical properties of composites containing various particle size distributions are presented in Section~\ref{distribution}. The conclusions and implications for applications are outlined in Section~\ref{concl}.

\section{Methods}\label{methods}

\textit{Materials:} Crushed silicon was purchased from Alfa Aesar. SEM imaging of the Si powder indicates that the received particles have an irregular shape (Figure~S1~\cite{SI}). Two fused silica powders, nano- and micron-sized silica powders, with a nominal particle size of 8 nm (from Sigma Aldrich) and 1~$\upmu$m (from Alfa Aesar) correspondingly were employed.

\textit{Composite compact preparation:} Reference silica and composite silica-silicon compacts were prepared by pulsed electric current sintering. Silicon particles were mixed with fused silica using Willy Bachofen Turbula Mixer for 2 hours. The reference silica powder or the mixed composite powder was placed into a mold with a 20 mm diameter. A 0.2 mm graphite foil was wrapped around the mold to ensure a tight fit. A heating rate of 100\degree~C/min and pressure of 50~MPa were applied during the entire experiment. A degassing step was used during the heating at 600\degree~C for 3 minutes to allow any gases in the powder to escape. All samples were sintered at a temperature of 1200\degree~C for 3 minutes for micro SiO$_2$ and 1 minute for nano SiO$_2$. The graphite paper was removed from the samples, and the samples were ground to a final surface quality using 1200-grit SiC paper followed by mechanical polishing using diamond paste with 6~$\upmu$m, 3~$\upmu$m, and 1~$\upmu$m particle sizes. The prepared samples were 20~mm in diameter and 3~mm in thickness. The density of the samples was measured by the Archimedes method using deionized water.

\textit{Characterization:} The microstructure of the samples was measured using X-ray diffraction (XRD) with a PANalytical Xpert Pro powder with a Co anode ($K~\alpha$) X-ray source. Scanning electron microscopy (SEM) images were obtained by TESCAN Mira 3 and Focused Ion Beam (FIB) Jeol JIB 4700. The UV-Vis-NIR diffuse reflectance spectrum was recorded in the wavelength range 250-2500 nm using a UV-Vis-NIR Agilent Cary 5000 equipped with diffuse reflectance accessories (integrating sphere). Static Light Scattering measurements were obtained with a Mastersizer 2000.

\textit{Radiative transport simulations:} The optical properties of the silica compacts embedded with silicon were calculated using a modified Monte Carlo method~\cite{tang2017plasmonically} adapted from Wang et al.~\cite{wang1995mcml}. A spherical particle with radius, $r$, was surrounded by a non-absorbing insulating medium with constant refractive index of 1.5 and irradiated by light, $\lambda =$ 250 to 2500 nm. The dielectric function of silicon was obtained from~\citet{palik1998handbook}. The efficiencies of scattering, $Q_{\rm{sca}}$, absorption, $Q_{\rm{abs}}$, and scattering asymmetry factor, $g$, of a single particle in an incident electromagnetic field were calculated using the full solutions of Lorenz-Mie theory and are described in the Supplemental Material~\cite{SI}.

Here, the microparticles were well-distributed in a non-absorbing insulating matrix with a volume fraction, $f$, up to 1\%. The compact thickness, $t$, was varied from 50~$\upmu$m to 3 mm, and was surrounded by a non-absorbing ambient medium. The results from theory were performed using the particle size distribution, $\rho (r)$, obtained from Static Light Scattering measurements. The ensemble averaged scattering and absorption coefficients, $\mu_{\rm{sca}}$ and $\mu_{\rm{abs}}$, of the compact are 
\begin{equation}\label{eqn:mus}
\mu_{\rm{sca,abs}} = \frac{3}{4} \sum_i \frac{f~ \rho(r_i)~Q_{\rm{sca,abs}}(r_i)}{r_i},
\end{equation}
\noindent where $Q_{\rm{sca,abs}}(r_i)$ is the scattering or absorption efficiency of a particle with radius, $r_i$, and $i$ is the binning index~\cite{merikallio2016computer}. The ensemble averaged particle asymmetry factor is
\begin{equation}
    g = \frac{\sum_i \rho(r_i)~Q_{\rm{sca}}(r_i) g(r_i)}{\sum_i \rho(r_i)~Q_{\rm{sca}}(r_i)}.
\end{equation}
\noindent The effective dielectric permittivity of the compact, $\epsilon_{\rm{eff}}$, from Maxwell Garnett Effective Medium Theory is~\cite{markel2016introduction}
\begin{equation}\label{eqn:eff}
\epsilon_{\rm{eff}} = \epsilon_{\rm{m}} + 3 f \epsilon_{\rm{m}} \frac{\epsilon_{\rm{p}} - \epsilon_{\rm{m}}}{\epsilon_{\rm{p}} + 2\epsilon_{\rm{m}} - f (\epsilon_{\rm{p}} - \epsilon_{\rm{m}})},    
\end{equation}
\noindent where $\epsilon_{\rm{p}}$ and $\epsilon_{\rm{m}}$ are the dielectric permittivities of the particle and medium components. The single particle scattering and absorption efficiencies and scattering anisotropy factor of different particle sizes are provided in the Supplemental Material~\cite{SI}.

The Monte Carlo method records the path and termination result of $10^7$ photons from an infinitesimally small beam normal to the compact surface. The reflected, absorbed, and transmitted photons are normalized to the total number of photons. A grid resolution of $dz = 0.1~\mu$m and $dr_{\textrm{grid}} = 5~\mu$m was used for the axial and radial directions, respectively. The total number of grid elements in the axial and angular directions were chosen to fit the compact thickness, $t$. The diffuse reflectance and transmittance go to zero as a function of the radius of the layer. The method for obtaining optical spectra of compacts containing particles with a distribution of sizes was verified for particles with a narrow distribution of sizes against compacts embedded with particles with uniform size~\cite{tang2017plasmonically}.

Additionally, the optical properties of the compacts were calculated using the Kubelka-Munk model~\cite{kubelka1948new,van1987tissue}. The Kubelka-Munk model describes the transport of light within a slab using the flux in the forward and reverse directions~\cite{murphy2006modified}. The reflectance of the slab representing the compact is
\begin{equation}\label{eqn:KM1}
    R_{\rm{KM}} = \frac{1}{a + b~\textrm{coth}~bSt}
\end{equation}
and the transmittance of the slab is
\begin{equation}\label{eqn:KM2}
    T_{\rm{KM}} = \frac{b~\textrm{csch}~bSt}{a + b~\textrm{coth}~bSt}
\end{equation}
where $t$ is the compact thickness, $a = 1 + K/S$, $b = (a^2 - 1)^{1/2}$, and the parameters $K$ and $S$ are obtained from the scattering and absorption coefficients as $K = 2\mu_{\rm{abs}}$ and $S = 3 (1 - g)\mu_{\rm{sca}}/4 - \mu_{\rm{abs}}/4$~\cite{slovick2015tailoring}. The absorption of the slab representing the compact from the Kubelka-Munk model is $A_{\rm{KM}} = 1 - R_{\rm{KM}} - T_{\rm{KM}}$. 

\section{Results and Discussion}
\subsection{Sintered compacts}\label{exp}

\begin{figure*}
\includegraphics[width=.9\textwidth]{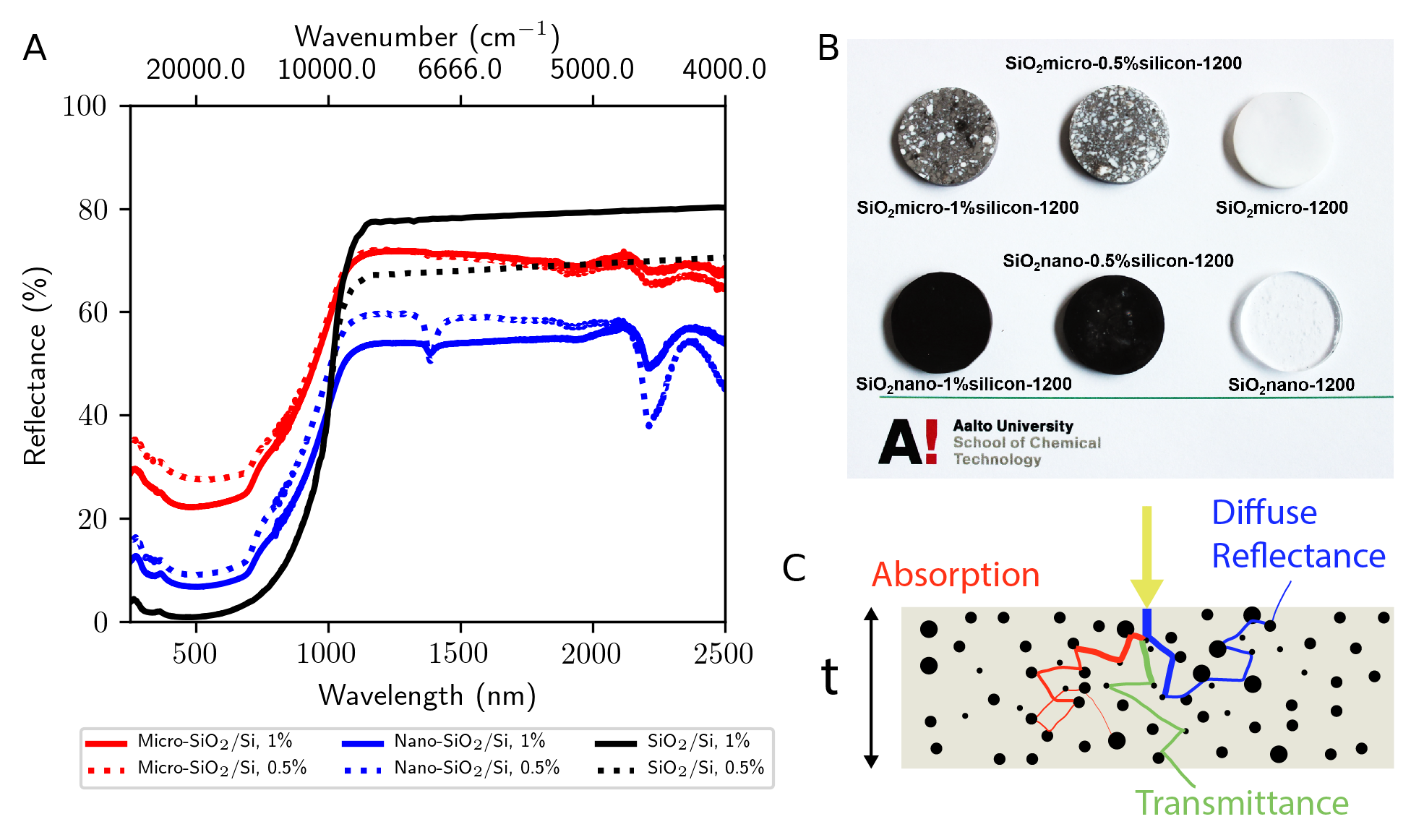}
\caption{A) Diffuse reflectance spectra of compacts consisting of Si particles embedded in SiO$_2$ consolidated from either micro- (red) or nano-SiO$_2$ powder (blue) and measured experimentally. The calculated diffuse reflectance spectra of the Si/SiO$_2$ composites are shown in black. The volume fraction of Si particles is either 0.5\% (dotted lines) or 1.0\% (solid lines). B) Compacts obtained by PECS at 1200\degree C. C) Schematic of the radiative transport within the specimen considered by the Monte Carlo method (see the details in Section~\ref{methods}). Specular reflectance is also considered at the front and rear of the compact.}
\label{fig:summary}
\end{figure*}

The compacts are sintered from either nano- or micron-sized silica powder and have different physical appearance, density, and optical properties (\textit{cf.} Figure~\ref{fig:summary}A). The distinct properties are attributed to the microstructure and post-sintering density. The specimens obtained from micro-silica powders have a more porous composite matrix than those consolidated from nano-silica. The porosity is observed in the relative density, diffuse reflectance, physical appearance, and the microstructure of the compact. 

\subsubsection{Relative density}
\begin{figure*}
\includegraphics[width=.9\textwidth]{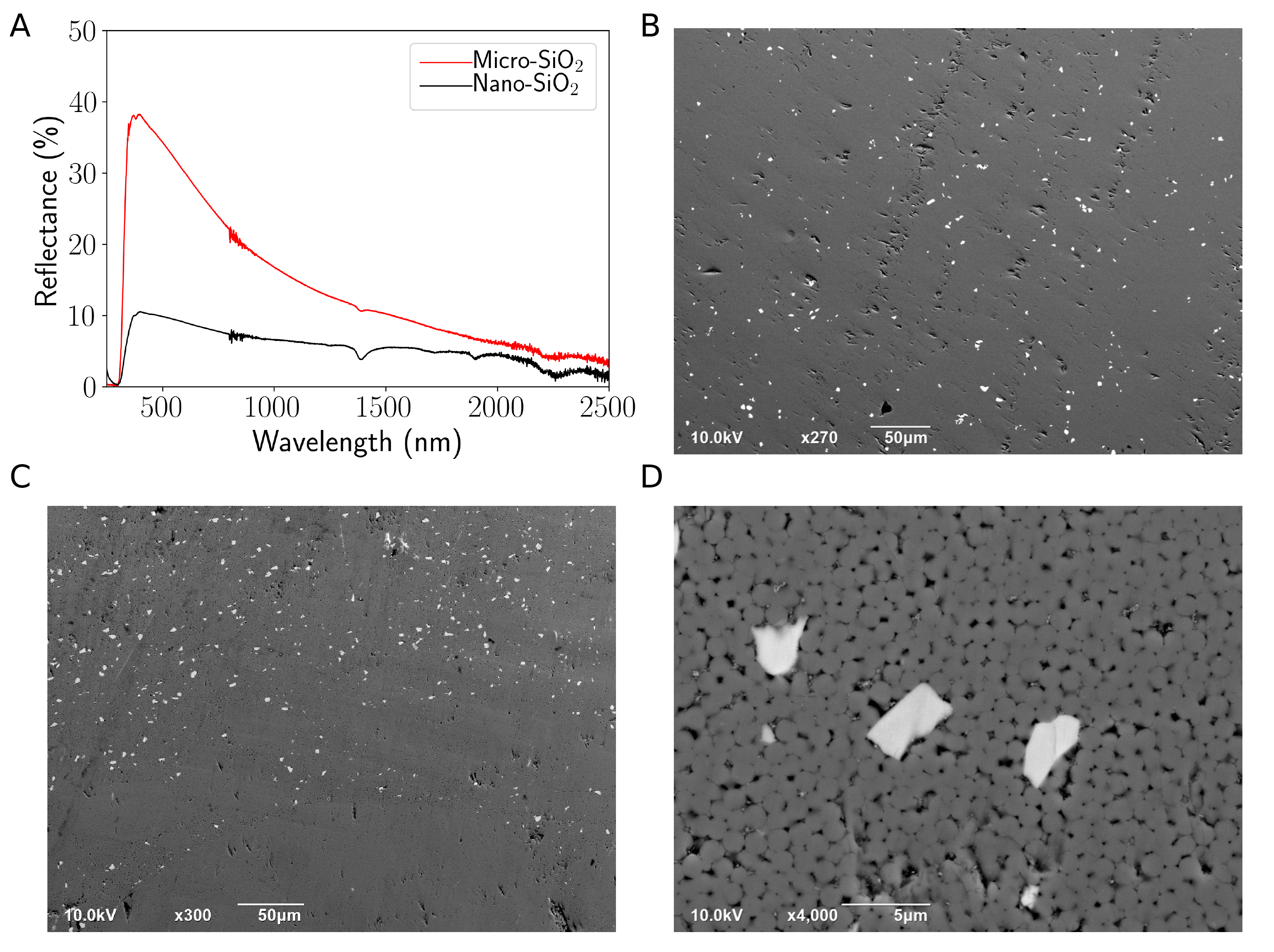}
\caption{A) Diffuse reflectance of nano- and micro-SiO$_2$ compacts without Si particles. Scanning electron micrographs of the surface of the (B) nano-silica and (C, D) micro-silica compacts embedded with Si particles at 1\%. The white specks are the silicon particles.}
\label{fig:exp}
\end{figure*}

The density of each composite layer relative to the density of non-porous silica (2.20 g/cm$^3$) indicates that the porosity differs for each compact (Table~\ref{table:density}). The compacts obtained from micro-silica powder have a density lower than those consolidated from nano-silica powder. The larger air-silica interface resulting from higher porosity serves to increase the scattering and consequently the diffuse reflectance of the micro-SiO$_2$ based compacts. This is indeed the case since compacts prepared from micro-silica powder consistently exhibit a larger diffuse reflectance than those prepared from nano-sized silica powder (Figure~\ref{fig:exp}A). 

\begin{table}
\caption{Density and relative density of silica-silicon samples prepared at 1200\degree C.}
\begin{tabular}{ |r|r|r| }
 \hline
Sample & Density (g/cm$^3$) & Relative Density (\%) \\ 
 \hline
Micro-SiO$_2$ / 0\% Si & 2.19 & 99.4 \\ 
/ 0.5\% Si & 1.98 & 90.1 \\ 
/ 1.0\% Si & 2.12 & 96.4 \\ 
\hline
Nano-SiO$_2$ / 0\% Si & 2.18 & 99.2 \\ 
 / 0.5\% Si & 2.15 & 97.8 \\ 
/ 1.0\% Si & 2.18 & 99.3 \\ 
 \hline
\end{tabular}
\label{table:density}
\end{table}

\subsubsection{Physical appearance}
Furthermore, the scattering in the visible region manifests in the physical appearance of the compacts (Figure~\ref{fig:summary}B). The compacts obtained from micro-silica powder are opaque while the nano-silica compact is transparent. The nano-silica based compacts containing Si microparticles are black, which indicates there is intense absorption in the visible region. The absorption is due to the generation of charge carriers from interband transitions at energies above the Si bandgap. On the other hand, the micro-silica/Si compacts are grayish in color due to the enhanced interfacial scattering within the more porous compact. 

\subsubsection{Microstructure}
Scanning Electron Microscopy of the compact surface visualizes and reveals the differences between the microstructure of the nano- and micro-silica compacts. The surface of the specimen consolidated from nano-silica powder is smooth and regular (Figure~\ref{fig:exp}B). The absence of large gaps and recesses suggests any pores in the nano-silica compact are likely nano-sized. The resulting packing density is very high and close to the bulk density of fused silica. Spatially resolved elemental analysis using energy dispersive X-ray spectroscopy (EDS) indicates that the white specks are silicon particles (Figure~S2~\cite{SI}). The particles are distributed evenly throughout the matrix. 

The microstructure is different in the compacts consolidated from micro-silica powder. The micro-silica compact without Si particles synthesized using PECS under identical conditions appears to be fully dense. When silicon particles are included, recesses and gaps form which increase the porosity of the compacts. The particles are unevenly distributed throughout the micro-silica matrix and are found predominately in randomly packed domains (Figures~\ref{fig:exp}C, D). The observed relative density of the micro-silica compacts (Table~\ref{table:density}) is consistent with a compact containing domains of cylindrical packing interspersed with random large holes (See Supplemental Material and Figure~S3 for details~\cite{SI}). The difference in sintering behavior of the Si-SiO$_2$ composite and pure micro-silica is likely to result from different interaction of the two kind of materials with PECS current, IR irradiation
imposed on the sample during PECS, and the large porosity introduced by Si particles.

The temperature and pressure during the PECS processing did not significantly change the size or shape of the silicon particles. When assumed to have a rectangular shape, the average particle size obtained by SEM image analysis was 2.4~$\times$~1.3~$\upmu$m and 2.5~$\times$~1.3~$\upmu$m for the nano- and micro-silica compacts, respectively. These sizes are slightly smaller than measured by Static Light Scattering (SLS) before the compact preparation in which the mode associated with the distribution of the particle diameters was 3.6~$\upmu$m.

Both nano- and micron-sized silica matrices of the consolidated samples have an amorphous structure. There are no crystalline peaks detected in the X-Ray diffractograms of the pure silica compacts (Figure~S4~\cite{SI}). Upon the inclusion of silicon particles to the mixture, crystalline peaks are detected in the X-Ray diffractograms. Samples with 0.5\% and 1.0\% microinclusion content exhibit diffraction peaks at 2$\theta$ = 32.42, 54.89, 65.69\degree~corresponding to the characteristic (111), (022), and (113) reflections of bulk Si, respectively. The maximum intensities of the lattice planes increase for the higher volume fraction of silicon.

\subsection{Comparison of experimental and calculated spectra}\label{spectra}
Next, the optical spectra of the silicon/silica compacts are examined. The reflectance has a minimum at short wavelengths (Figure~\ref{fig:summary}A). Beginning at about 500 nm, the reflectance steadily increases to a plateau at about 1100 nm. The edge of the plateau at shorter wavelength corresponds to the bandgap of Si (1100 nm, 1.1 eV) and is expected behavior as charge carriers are generated due to interband transitions at energies beyond the absorption band edge. Monte Carlo simulations provide information about the interplay between absorption and scattering processes within the compacts of varying thickness and volume fraction. The spectra are calculated for a non-porous compact with well-dispersed spherical particles, which results in some differences with the experimental spectra. The analysis is divided into three spectral regions demarcated by (region I) the reflectance plateau at $\lambda >$ 1100 nm in which there is strong reflectance and negligible absorption, (region II) the absorption region at $\lambda <$ 750 nm in which the absorption dominates, and (region III) a transition region between $\lambda =$ 750 nm and 1100 nm which is characterized by a weakening absorption and growing reflectance. 

\begin{figure*}
\includegraphics[width=\textwidth]{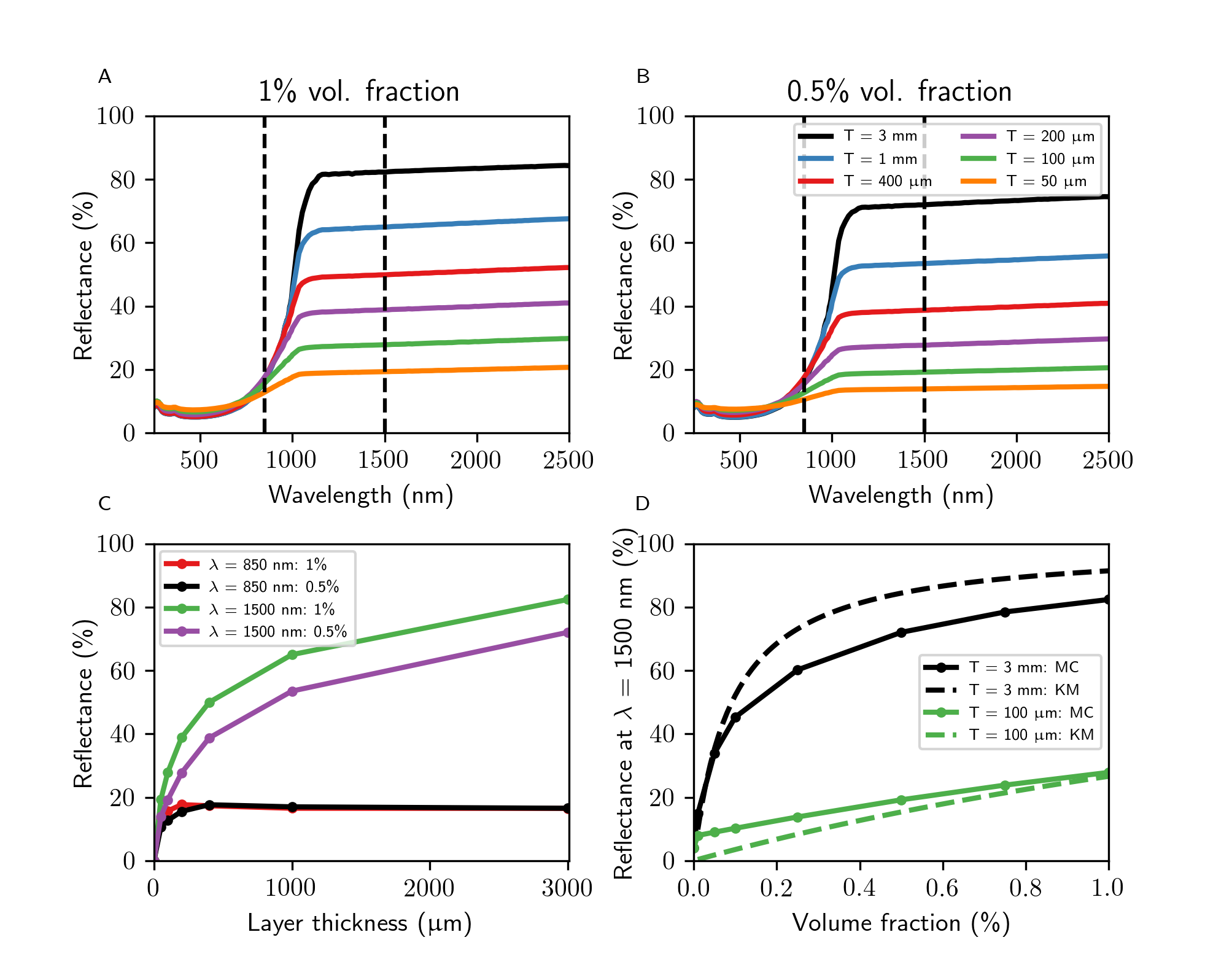}
\caption{Calculated reflectance spectra of silica-silicon compacts of thickness, $t$ = 50 $\upmu$m, 100 $\upmu$m, 200 $\upmu$m, 400 $\upmu$m, 1 mm, and 3 mm with A) 1\% volume fraction B) 0.5\% volume fraction. C) Total reflectance at $\lambda$ = 850 nm and 1500 nm for different compact thicknesses. D) Total reflectance at $\lambda$ = 1500 nm for different volume fractions and calculated with either Monte Carlo (MC, solid lines) or the Kubelka-Munk model (KM, dashed lines). See Eqn.~(\ref{eqn:KM1})~and~(\ref{eqn:KM2}) in Section~\ref{methods} for details of the KM model. The solid lines in C) and D) are drawn to guide the eye.}
\label{fig:sim}
\end{figure*}

\subsubsection{Region I.}
First, in the reflectance plateau (region I), the radiative transport within the silica compacts is dominated by scattering from the embedded Si microparticles. The calculated scattering coefficient, $\mu_{\rm{sca}}$, is large and the absorption coefficient, $\mu_{\rm{abs}}$, is negligible (109.4 cm$^{-1}$ and $3 \times 10^{-3}$ cm$^{-1}$ at 1500 nm, respectively at 1\%). The scattering anisotropy, $g$, is approximately 0.5, and yet despite the scattering being moderately forward, the compacts are strongly reflective because the direction of the incident radiation is quickly randomized due to scattering from the randomly dispersed Si microparticles~\cite{tang2017plasmonically}. 

The effective medium is derived from independent scattering events from single spherical particles. The scattering energies of an irregular particle are expected to be slightly different than a spherical particle~\cite{tzarouchis2016shape}. The experimentally measured reflectance might be less than the calculated reflectance due to particle clustering and irregular morphology in the manufactured compacts. We note that it is not uncommon to model scattering from a distribution of irregularly shaped particles using ensembles of regular model particles such as spheres, spheroids, or ellipsoids~\cite{merikallio2016computer}. 

The theoretical spectra show the magnitude of the reflectance plateau increases when either the compact thickness or particle volume fraction is increased (Figures~\ref{fig:sim}A, B). The dependence of the total reflectance at $\lambda$ = 1500 nm on the thickness is shown in Figure~\ref{fig:sim}C. By elongating the physical path length, \textit{i.e.} the compact thickness, it becomes more probable that the photon direction has reversed from either a back scattering event or enough diffuse scattering events. The scattering coefficient is directly proportional to the volume fraction of particles (see Eqn.~(\ref{eqn:mus}) in the Methods section below), and the magnitude of the plateau is larger in 1\% compacts than in 0.5\% due to the larger scattering coefficient. The effect of volume fraction on the magnitude of the reflectance plateau is examined in Figure~\ref{fig:sim}D. The non-monotonic increase of the reflectance is more pronounced in the thicker compacts. The non-monotonic behavior is consistent with the results from the Kubelka-Munk model. Due to the two-flux simplification and the absence of specular reflectance in the Kubelka-Munk model~\cite{vargas1997applicability}, the intensity of reflectance is stronger in thicker coating but weaker in the thin coating compared to the results from the Monte Carlo model. See Eqn.~(\ref{eqn:KM1})~and~(\ref{eqn:KM2}) in Section~\ref{methods} for details. The Monte Carlo method is more accurate because it includes specular reflectance at the front and rear of the composite.

The diffuse reflectance was not detected beyond 2.5~$\upmu$m experimentally. The computational modelling predicts broad and intense scattering until about 10~$\upmu$m, and absorption modes are absent in the calculated reflectance spectra which assume a non-absorbing medium with a refractive index of 1.5. Beginning around at 4.5~$\upmu$m, there is expected to be a pronounced effect from the absorption from the Si-O stretching modes within the medium. Absorption by oxides have previously been considered in the shell of semiconductor particles coated with an oxide layer~\cite{conley2018plasmonically,conley2020directing}. Indeed, FTIR shows strong absorption in this region (Figure~S5~\cite{SI}). Other absorption modes within the silica medium produce dips in the reflectance plateau at 1.39 and 2.20~$\upmu$m. These features are also present in the compacts without silicon (Figure~\ref{fig:exp}A). 

\begin{figure*}
\includegraphics[width=\textwidth]{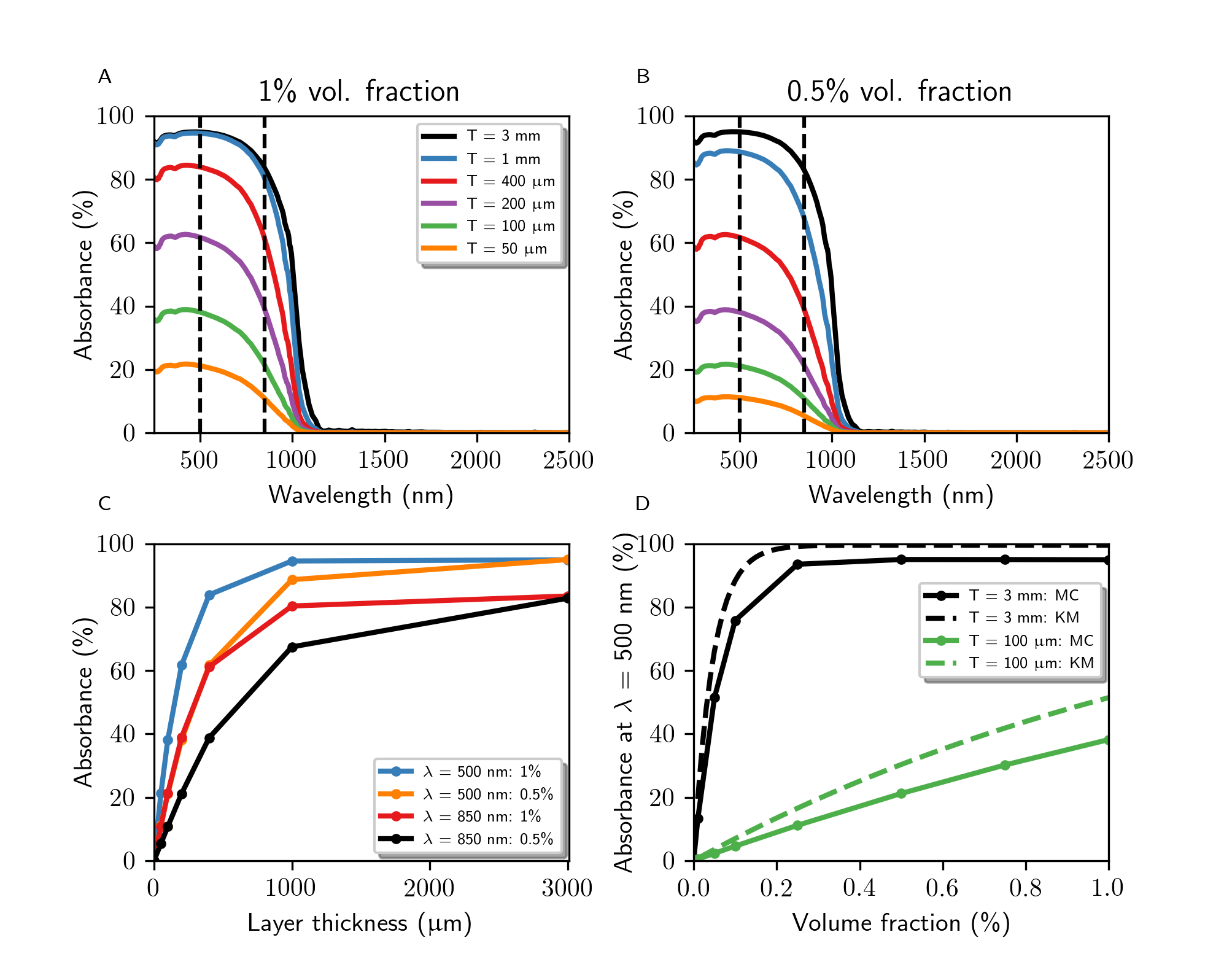}
\caption{Calculated absorption spectra of silica-silicon compacts of thickness, $t$ = 50 $\upmu$m, 100 $\upmu$m, 200 $\upmu$m, 400 $\upmu$m, 1 mm, and 3 mm with A) 1\% volume fraction B) 0.5\% volume fraction. C) Absorption at $\lambda$ = 500 nm and 850 nm for different compact thicknesses. D) Absorption at $\lambda$ = 500 nm for different volume fractions and calculated with either Monte Carlo (MC, solid lines) or the Kubelka-Munk model (KM, dashed lines). See Eqn.~(\ref{eqn:KM1})~and~(\ref{eqn:KM2}) in Section~\ref{methods} for details of the KM model. The solid lines in C) and D) are drawn to guide the eye.} 
\label{fig:sim_abs}
\end{figure*}

\subsubsection{Region II.}
Second, at short wavelengths in region II, absorption by the silicon particles becomes significant enough to be a competitive transport mechanism and influence the optical behavior. Intense absorption in the visible region is observed (Figure~\ref{fig:sim_abs}) as expected from the black color of the compacts due to the interband transitions at energies above the Si bandgap. The total absorption increases with the thickness of the silica compact or the volume fraction of the embedded Si particles. The calculated absorption at 500 nm is examined more closely in Figure~\ref{fig:sim_abs}C, D, and the absorption plateau is reached for a 1 mm thick compact at 1\% volume fraction of Si particles or 0.2\% for a 3~mm thick compact. The compacts do not fully absorb due to the specular reflectance at the front air-composite interface, which accounts for about 4\% of the intensity. The Kubelka-Munk model does not include specular reflectance and the thick compacts completely absorb the incident radiation for volume fractions greater than 0.25\%.

The scattering enhances the absorption by elongating the optical path length and enabling more opportunities for absorption. Despite having a non-zero scattering coefficient, both the experimentally measured and calculated reflectance decrease in compacts that are either thicker or contain more particles. It is not necessary for $\mu_{\rm{abs}}$ to be larger than $\mu_{\rm{sca}}$ for the absorption processes to dominate the transport. In fact $\mu_{\rm{sca}}$ is as much as to 3.5 times larger than $\mu_{\rm{abs}}$ in this region, and yet the reflectance is less than 9\% while the absorption is greater than 20\% for compacts greater than 50 $\upmu$m thick ($\mu_{\rm{abs}}$ = 36.2 cm$^{-1}$ and $\mu_{\rm{sca}}$ = 64.0 cm$^{-1}$ at 500 nm and 1\% Si). The imbalance is due to the nature of the transport processes. To reflect light, the photon direction needs to both reverse and avoid absorption, but absorption can terminate the photon propagation anywhere in the compact.

\subsubsection{Region III.}
Finally, in region III between about 750 and 1100 nm, the calculated $\mu_{\rm{abs}}$ is low but not insignificant. The optical properties transition between the highly reflective region in which the scattering processes dominate ($\lambda >$ 1100 nm) and the highly absorbing region in which the photons are more likely to be absorbed thereby culling transport ($\lambda <$ 750 nm). At $\lambda$ = 850~nm, the reflectance and absorption saturate to a maximum of 16\% and 84\% in Figures~\ref{fig:sim}C and~\ref{fig:sim_abs}C, respectively. Notably, the plateau of maximum reflectance is reached in thinner compacts ($t$ = 400 $\upmu$m) than the plateau of maximum absorption ($t$ = 3 mm). At this wavelength, compacts thicker than 400 $\upmu$m increase the absorption but not the reflectance. The diffuse scattering which results in reflectance occurs in the front 400~$\upmu$m of the compact.

\subsection{Effect of particle size distribution}\label{distribution}

\begin{figure*}
\includegraphics[width=\textwidth]{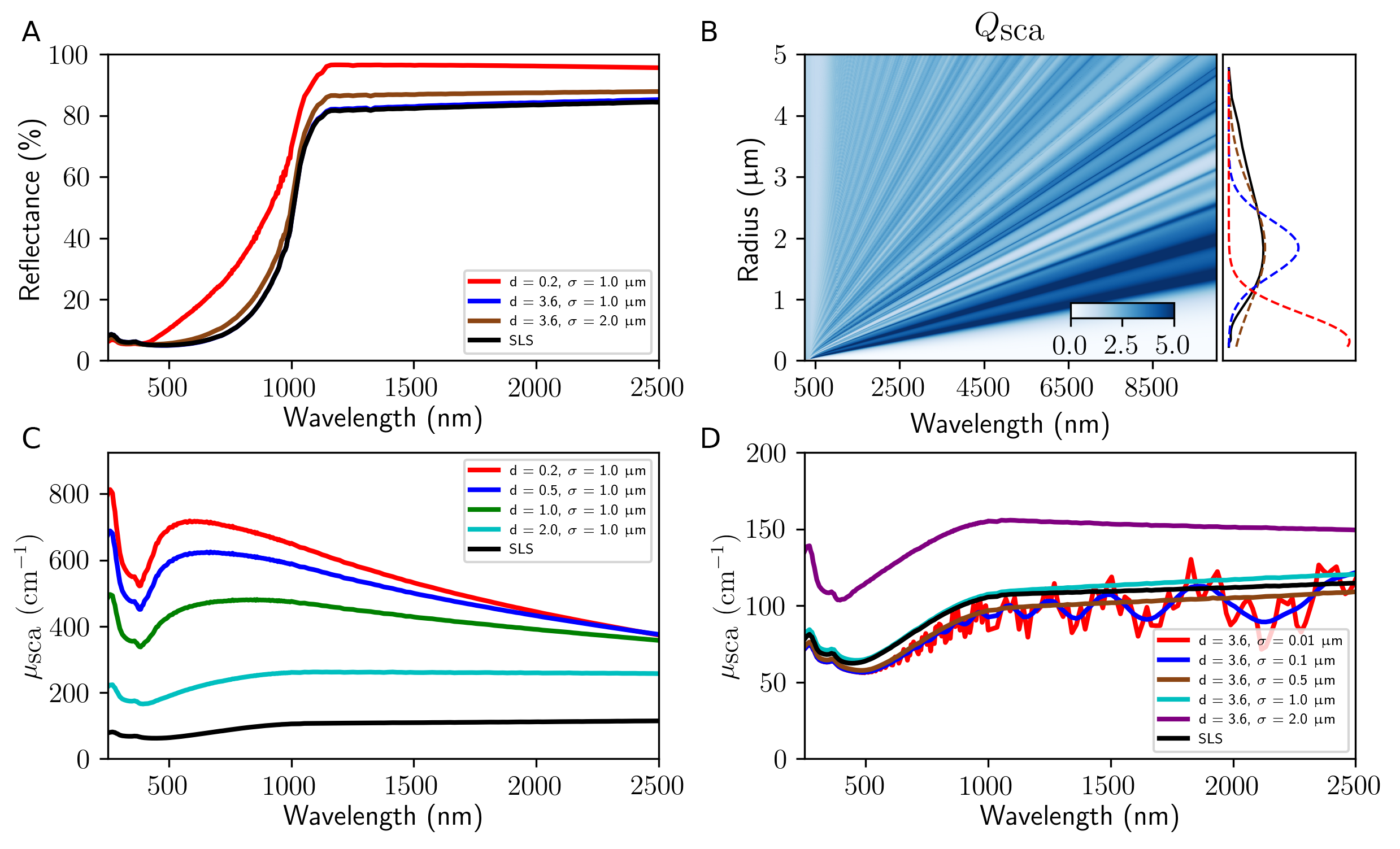}
\caption{A) Calculated reflectance spectra of compacts containing particles with different size distributions. B) Single particle scattering absorption efficiencies for particles with radius up to 5~$\upmu$m. Also shown are the size distributions of Si particles obtained by Static Light Scattering and truncated normal distributions with an average particle diameter, $d$, of 3.6 or 0.2~$\upmu$m and standard deviation, $\sigma$, of 2.0 or 1.0~$\upmu$m. Calculated (C, B) scattering coefficients of compacts with 1\% volume fraction of silicon. The scattering coefficients of the experimentally measured distribution are compared with the theoretical size distributions with (C) mean diameter equal 3.6 $\upmu$m and standard deviation, $\sigma$, varying from 0.01 to 2.0 $\upmu$m and (D) the mean particle diameter is shifted to 1.0 $\upmu$m with a standard deviation of 1.0 $\upmu$m.}
\label{fig:size_dist}
\end{figure*}

The scattering resonance energies vary with the particle size and its distribution, and the optical properties of compacts containing other size distributions are compared with compacts containing the experimentally measured size distribution (Figure~\ref{fig:size_dist}A). The theoretical particle sizes are truncated normal distributions and provided alongside the single particle scattering efficiencies for the particle sizes in Figure~\ref{fig:size_dist}B. The reflectance is strengthened by either decreasing the average particle size or including more small particles. Decreasing the average particle size has a more drastic increase on the reflectance than changing the standard deviation and the consequences are discussed for compacts containing 1\% Si particles below.

The reflectance increases in compacts containing a smaller average particle size than the experimentally measured mode diameter (3.6~$\upmu$m). For a fixed volume fraction, the larger particles occupy volume without significantly enhancing the scattering or absorption coefficients employed in the Monte Carlo modeling of radiation transport. See Section~\ref{methods} for details. Thus, the particle number density increases with a lower average size. This trend is illustrated in the significantly larger scattering coefficients for particles with a smaller average particle diameter than 3.6~$\upmu$m (Figure~\ref{fig:size_dist}C). The more intense scattering coefficient at shorter wavelengths changes the reflectance profile in the compacts containing smaller particles. While the absorption coefficient is also increased for smaller average particle sizes (Figures~S6~and~S7~\cite{SI}), the absorbance is overcome by the strong scattering.

Next, the standard deviation is varied for distributions with an average diameter equal to the mode experimentally measured particle diameter (Figure~\ref{fig:size_dist}D). As the distribution broadens, the ripple structure of the scattering coefficients flattens and the magnitude of $\mu_{\rm{sca}}$ increases to 150~cm$^{-1}$ at 1500 nm for the distribution with $\sigma = 2~\upmu$m. Although this broad distribution contains more large particles, their parasitic effects are outweighed by the advantage of including effective small particles which strengthen the scattering. Similarly, the magnitude of the absorption coefficient increases as the particle size distribution broadens (Figure~S7~\cite{SI}).

\section{Conclusions}\label{concl}
PECS is a facile, straight-forward approach to manufacture composites. Unlike other manufacturing techniques, PECS has the potential to produce NIR reflective compacts at large-scale and with a cost-effective method. Silica-silicon particle composite compacts were synthesized using PECS from either nano-silica or micro-silica powder and their optical properties compared with the results from calculations. 

Incorporating a small volume fraction of micro-Si particles changes the composites from short wavelength reflectors (Figure~\ref{fig:exp}A) to long wavelength reflectors (Figure~\ref{fig:summary}A). Diffuse reflectance of up to 72\% can be achieved with silicon particle volume fractions of 0.5\% and 1\% in 3~mm thick compacts. The specimens obtained from nano-silica powders had a denser and less porous composite matrix than those consolidated from micro-silica. The rapid processing time of PECS limits the undesirable transformations such as crystallization, oxidation, and grain growth. 

The broadband NIR reflectance was obtained with a wide size distribution of microparticles. The size distribution affects the reflectance intensity and profile, and composites containing larger particle number densities were shown to have stronger reflectance. The spectra calculated using Monte Carlo and Kubelka-Munk methods predicted that a significant increase in reflectance can be obtained by lowering the average particle size as opposed to a narrower particle size distribution. However, composites with a narrow size distribution that can selectively reflect in a given wavelength region can be exploited for applications in gradient heat flux sensors or detectors. 

Given the low dissipation in semiconductor microinclusions, these composites can be used to control the radiative transfer of heat in devices at high temperature. Our ongoing and future work is focused on the development of composites containing bandgap engineered semiconductor particles in multi-layered systems with a graded distribution in a wide range of host materials to enhance radiative transfer mechanisms in targeted applications. 

\begin{acknowledgments}
This work was performed as part of the Academy of Finland project 314488 and QTF Centre of Excellence program (312298, KMC and TAN). We acknowledge the provision of facilities and technical support by Aalto University at OtaNano - Nanomicroscopy Center (Aalto-NMC); computational resources provided by CSC -- IT Center for Science (Finland) and by the Aalto Science-IT project (Aalto University School of Science); Natural Sciences and Engineering Research Council (NSERC) of Canada (MK); Canada Research Chairs Program (MK); and Compute Canada (www.computecanada.ca).

\end{acknowledgments}

\bibliography{main_arxiv.bbl}

\begin{thebibliography}{34}%
\makeatletter
\providecommand \@ifxundefined [1]{%
 \@ifx{#1\undefined}
}%
\providecommand \@ifnum [1]{%
 \ifnum #1\expandafter \@firstoftwo
 \else \expandafter \@secondoftwo
 \fi
}%
\providecommand \@ifx [1]{%
 \ifx #1\expandafter \@firstoftwo
 \else \expandafter \@secondoftwo
 \fi
}%
\providecommand \natexlab [1]{#1}%
\providecommand \enquote  [1]{``#1''}%
\providecommand \bibnamefont  [1]{#1}%
\providecommand \bibfnamefont [1]{#1}%
\providecommand \citenamefont [1]{#1}%
\providecommand \href@noop [0]{\@secondoftwo}%
\providecommand \href [0]{\begingroup \@sanitize@url \@href}%
\providecommand \@href[1]{\@@startlink{#1}\@@href}%
\providecommand \@@href[1]{\endgroup#1\@@endlink}%
\providecommand \@sanitize@url [0]{\catcode `\\12\catcode `\$12\catcode
  `\&12\catcode `\#12\catcode `\^12\catcode `\_12\catcode `\%12\relax}%
\providecommand \@@startlink[1]{}%
\providecommand \@@endlink[0]{}%
\providecommand \url  [0]{\begingroup\@sanitize@url \@url }%
\providecommand \@url [1]{\endgroup\@href {#1}{\urlprefix }}%
\providecommand \urlprefix  [0]{URL }%
\providecommand \Eprint [0]{\href }%
\providecommand \doibase [0]{http://dx.doi.org/}%
\providecommand \selectlanguage [0]{\@gobble}%
\providecommand \bibinfo  [0]{\@secondoftwo}%
\providecommand \bibfield  [0]{\@secondoftwo}%
\providecommand \translation [1]{[#1]}%
\providecommand \BibitemOpen [0]{}%
\providecommand \bibitemStop [0]{}%
\providecommand \bibitemNoStop [0]{.\EOS\space}%
\providecommand \EOS [0]{\spacefactor3000\relax}%
\providecommand \BibitemShut  [1]{\csname bibitem#1\endcsname}%
\let\auto@bib@innerbib\@empty
\bibitem [{\citenamefont {Kabashin}\ \emph {et~al.}(2009)\citenamefont
  {Kabashin}, \citenamefont {Evans}, \citenamefont {Pastkovsky}, \citenamefont
  {Hendren}, \citenamefont {Wurtz}, \citenamefont {Atkinson}, \citenamefont
  {Pollard}, \citenamefont {Podolskiy},\ and\ \citenamefont
  {Zayats}}]{kabashin2009plasmonic}%
  \BibitemOpen
  \bibfield  {author} {\bibinfo {author} {\bibfnamefont {A.}~\bibnamefont
  {Kabashin}}, \bibinfo {author} {\bibfnamefont {P.}~\bibnamefont {Evans}},
  \bibinfo {author} {\bibfnamefont {S.}~\bibnamefont {Pastkovsky}}, \bibinfo
  {author} {\bibfnamefont {W.}~\bibnamefont {Hendren}}, \bibinfo {author}
  {\bibfnamefont {G.}~\bibnamefont {Wurtz}}, \bibinfo {author} {\bibfnamefont
  {R.}~\bibnamefont {Atkinson}}, \bibinfo {author} {\bibfnamefont
  {R.}~\bibnamefont {Pollard}}, \bibinfo {author} {\bibfnamefont
  {V.}~\bibnamefont {Podolskiy}}, \ and\ \bibinfo {author} {\bibfnamefont
  {A.}~\bibnamefont {Zayats}},\ }\href@noop {} {\bibfield  {journal} {\bibinfo
  {journal} {Nat. Mater.}\ }\textbf {\bibinfo {volume} {8}},\ \bibinfo {pages}
  {867} (\bibinfo {year} {2009})}\BibitemShut {NoStop}%
\bibitem [{\citenamefont {J\"{o}nsson}\ \emph {et~al.}(2017)\citenamefont
  {J\"{o}nsson}, \citenamefont {Tordera}, \citenamefont {Pakizeh},
  \citenamefont {Jaysankar}, \citenamefont {Miljkovic}, \citenamefont {Tong},
  \citenamefont {Jonsson},\ and\ \citenamefont {Dmitriev}}]{jonsson2017solar}%
  \BibitemOpen
  \bibfield  {author} {\bibinfo {author} {\bibfnamefont {G.}~\bibnamefont
  {J\"{o}nsson}}, \bibinfo {author} {\bibfnamefont {D.}~\bibnamefont
  {Tordera}}, \bibinfo {author} {\bibfnamefont {T.}~\bibnamefont {Pakizeh}},
  \bibinfo {author} {\bibfnamefont {M.}~\bibnamefont {Jaysankar}}, \bibinfo
  {author} {\bibfnamefont {V.}~\bibnamefont {Miljkovic}}, \bibinfo {author}
  {\bibfnamefont {L.}~\bibnamefont {Tong}}, \bibinfo {author} {\bibfnamefont
  {M.~P.}\ \bibnamefont {Jonsson}}, \ and\ \bibinfo {author} {\bibfnamefont
  {A.}~\bibnamefont {Dmitriev}},\ }\href@noop {} {\bibfield  {journal}
  {\bibinfo  {journal} {Nano Lett.}\ }\textbf {\bibinfo {volume} {17}},\
  \bibinfo {pages} {6766} (\bibinfo {year} {2017})}\BibitemShut {NoStop}%
\bibitem [{\citenamefont {Ma}\ \emph {et~al.}(2018)\citenamefont {Ma},
  \citenamefont {Zagar}, \citenamefont {Klemme}, \citenamefont {Sikdar},
  \citenamefont {Velleman}, \citenamefont {Montelongo}, \citenamefont {Oh},
  \citenamefont {Kucernak}, \citenamefont {Edel},\ and\ \citenamefont
  {Kornyshev}}]{ma2018tunable}%
  \BibitemOpen
  \bibfield  {author} {\bibinfo {author} {\bibfnamefont {Y.}~\bibnamefont
  {Ma}}, \bibinfo {author} {\bibfnamefont {C.}~\bibnamefont {Zagar}}, \bibinfo
  {author} {\bibfnamefont {D.~J.}\ \bibnamefont {Klemme}}, \bibinfo {author}
  {\bibfnamefont {D.}~\bibnamefont {Sikdar}}, \bibinfo {author} {\bibfnamefont
  {L.}~\bibnamefont {Velleman}}, \bibinfo {author} {\bibfnamefont
  {Y.}~\bibnamefont {Montelongo}}, \bibinfo {author} {\bibfnamefont {S.-H.}\
  \bibnamefont {Oh}}, \bibinfo {author} {\bibfnamefont {A.~R.}\ \bibnamefont
  {Kucernak}}, \bibinfo {author} {\bibfnamefont {J.~B.}\ \bibnamefont {Edel}},
  \ and\ \bibinfo {author} {\bibfnamefont {A.~A.}\ \bibnamefont {Kornyshev}},\
  }\href@noop {} {\bibfield  {journal} {\bibinfo  {journal} {ACS Photonics}\
  }\textbf {\bibinfo {volume} {5}},\ \bibinfo {pages} {4604} (\bibinfo {year}
  {2018})}\BibitemShut {NoStop}%
\bibitem [{\citenamefont {Sousa-Castillo}\ \emph {et~al.}(2017)\citenamefont
  {Sousa-Castillo}, \citenamefont {Ameneiro-Prieto}, \citenamefont
  {Comesa{\~n}a-Hermo}, \citenamefont {Yu}, \citenamefont
  {Vila-Fungueiri{\~n}o}, \citenamefont {P{\'e}rez-Lorenzo}, \citenamefont
  {Rivadulla}, \citenamefont {de~Abajo},\ and\ \citenamefont
  {Correa-Duarte}}]{sousa2017hybrid}%
  \BibitemOpen
  \bibfield  {author} {\bibinfo {author} {\bibfnamefont {A.}~\bibnamefont
  {Sousa-Castillo}}, \bibinfo {author} {\bibfnamefont {{\'O}.}~\bibnamefont
  {Ameneiro-Prieto}}, \bibinfo {author} {\bibfnamefont {M.}~\bibnamefont
  {Comesa{\~n}a-Hermo}}, \bibinfo {author} {\bibfnamefont {R.}~\bibnamefont
  {Yu}}, \bibinfo {author} {\bibfnamefont {J.~M.}\ \bibnamefont
  {Vila-Fungueiri{\~n}o}}, \bibinfo {author} {\bibfnamefont {M.}~\bibnamefont
  {P{\'e}rez-Lorenzo}}, \bibinfo {author} {\bibfnamefont {F.}~\bibnamefont
  {Rivadulla}}, \bibinfo {author} {\bibfnamefont {F.~J.~G.}\ \bibnamefont
  {de~Abajo}}, \ and\ \bibinfo {author} {\bibfnamefont {M.~A.}\ \bibnamefont
  {Correa-Duarte}},\ }\href@noop {} {\bibfield  {journal} {\bibinfo  {journal}
  {Nano Energy}\ }\textbf {\bibinfo {volume} {37}},\ \bibinfo {pages} {118}
  (\bibinfo {year} {2017})}\BibitemShut {NoStop}%
\bibitem [{\citenamefont {Mityakov}\ \emph {et~al.}(2012)\citenamefont
  {Mityakov}, \citenamefont {Sapozhnikov}, \citenamefont {Mityakov},
  \citenamefont {Snarskii}, \citenamefont {Zhenirovsky},\ and\ \citenamefont
  {Pyrh{\"o}nen}}]{mityakov2012gradient}%
  \BibitemOpen
  \bibfield  {author} {\bibinfo {author} {\bibfnamefont {A.~V.}\ \bibnamefont
  {Mityakov}}, \bibinfo {author} {\bibfnamefont {S.~Z.}\ \bibnamefont
  {Sapozhnikov}}, \bibinfo {author} {\bibfnamefont {V.~Y.}\ \bibnamefont
  {Mityakov}}, \bibinfo {author} {\bibfnamefont {A.~A.}\ \bibnamefont
  {Snarskii}}, \bibinfo {author} {\bibfnamefont {M.~I.}\ \bibnamefont
  {Zhenirovsky}}, \ and\ \bibinfo {author} {\bibfnamefont {J.~J.}\ \bibnamefont
  {Pyrh{\"o}nen}},\ }\href@noop {} {\bibfield  {journal} {\bibinfo  {journal}
  {Sens. Actuator A-Phys.}\ }\textbf {\bibinfo {volume} {176}},\ \bibinfo
  {pages} {1} (\bibinfo {year} {2012})}\BibitemShut {NoStop}%
\bibitem [{\citenamefont {Siebentritt}(2017)}]{siebentritt2017chalcopyrite}%
  \BibitemOpen
  \bibfield  {author} {\bibinfo {author} {\bibfnamefont {S.}~\bibnamefont
  {Siebentritt}},\ }\href@noop {} {\bibfield  {journal} {\bibinfo  {journal}
  {Curr. Opin. Green Sustain. Chem.}\ }\textbf {\bibinfo {volume} {4}},\
  \bibinfo {pages} {1} (\bibinfo {year} {2017})}\BibitemShut {NoStop}%
\bibitem [{\citenamefont {Yin}\ \emph {et~al.}(2015)\citenamefont {Yin},
  \citenamefont {Steigert}, \citenamefont {Andrae}, \citenamefont {Goebelt},
  \citenamefont {Latzel}, \citenamefont {Manley}, \citenamefont {Lauermann},
  \citenamefont {Christiansen},\ and\ \citenamefont
  {Schmid}}]{yin2015integration}%
  \BibitemOpen
  \bibfield  {author} {\bibinfo {author} {\bibfnamefont {G.}~\bibnamefont
  {Yin}}, \bibinfo {author} {\bibfnamefont {A.}~\bibnamefont {Steigert}},
  \bibinfo {author} {\bibfnamefont {P.}~\bibnamefont {Andrae}}, \bibinfo
  {author} {\bibfnamefont {M.}~\bibnamefont {Goebelt}}, \bibinfo {author}
  {\bibfnamefont {M.}~\bibnamefont {Latzel}}, \bibinfo {author} {\bibfnamefont
  {P.}~\bibnamefont {Manley}}, \bibinfo {author} {\bibfnamefont
  {I.}~\bibnamefont {Lauermann}}, \bibinfo {author} {\bibfnamefont
  {S.}~\bibnamefont {Christiansen}}, \ and\ \bibinfo {author} {\bibfnamefont
  {M.}~\bibnamefont {Schmid}},\ }\href@noop {} {\bibfield  {journal} {\bibinfo
  {journal} {Appl. Surf. Sci.}\ }\textbf {\bibinfo {volume} {355}},\ \bibinfo
  {pages} {800} (\bibinfo {year} {2015})}\BibitemShut {NoStop}%
\bibitem [{\citenamefont {van Lare}\ \emph {et~al.}(2015)\citenamefont {van
  Lare}, \citenamefont {Yin}, \citenamefont {Polman},\ and\ \citenamefont
  {Schmid}}]{van2015light}%
  \BibitemOpen
  \bibfield  {author} {\bibinfo {author} {\bibfnamefont {C.}~\bibnamefont {van
  Lare}}, \bibinfo {author} {\bibfnamefont {G.}~\bibnamefont {Yin}}, \bibinfo
  {author} {\bibfnamefont {A.}~\bibnamefont {Polman}}, \ and\ \bibinfo {author}
  {\bibfnamefont {M.}~\bibnamefont {Schmid}},\ }\href@noop {} {\bibfield
  {journal} {\bibinfo  {journal} {ACS Nano}\ }\textbf {\bibinfo {volume} {9}},\
  \bibinfo {pages} {9603} (\bibinfo {year} {2015})}\BibitemShut {NoStop}%
\bibitem [{\citenamefont {Baffou}\ and\ \citenamefont
  {Quidant}(2013)}]{baffou2013thermo}%
  \BibitemOpen
  \bibfield  {author} {\bibinfo {author} {\bibfnamefont {G.}~\bibnamefont
  {Baffou}}\ and\ \bibinfo {author} {\bibfnamefont {R.}~\bibnamefont
  {Quidant}},\ }\href@noop {} {\bibfield  {journal} {\bibinfo  {journal} {Laser
  Photonics Rev.}\ }\textbf {\bibinfo {volume} {7}},\ \bibinfo {pages} {171}
  (\bibinfo {year} {2013})}\BibitemShut {NoStop}%
\bibitem [{\citenamefont {Thakore}\ \emph {et~al.}(2019)\citenamefont
  {Thakore}, \citenamefont {Tang}, \citenamefont {Conley}, \citenamefont
  {Ala-Nissila},\ and\ \citenamefont {Karttunen}}]{thakore2019thermoplasmonic}%
  \BibitemOpen
  \bibfield  {author} {\bibinfo {author} {\bibfnamefont {V.}~\bibnamefont
  {Thakore}}, \bibinfo {author} {\bibfnamefont {J.}~\bibnamefont {Tang}},
  \bibinfo {author} {\bibfnamefont {K.}~\bibnamefont {Conley}}, \bibinfo
  {author} {\bibfnamefont {T.}~\bibnamefont {Ala-Nissila}}, \ and\ \bibinfo
  {author} {\bibfnamefont {M.}~\bibnamefont {Karttunen}},\ }\href@noop {}
  {\bibfield  {journal} {\bibinfo  {journal} {Adv. Theory Simul.}\ }\textbf
  {\bibinfo {volume} {2}},\ \bibinfo {pages} {1800100} (\bibinfo {year}
  {2019})}\BibitemShut {NoStop}%
\bibitem [{\citenamefont {Conley}\ \emph {et~al.}(2020)\citenamefont {Conley},
  \citenamefont {Thakore}, \citenamefont {Seyedheydari}, \citenamefont
  {Karttunen},\ and\ \citenamefont {Ala-Nissila}}]{conley2020directing}%
  \BibitemOpen
  \bibfield  {author} {\bibinfo {author} {\bibfnamefont {K.~M.}\ \bibnamefont
  {Conley}}, \bibinfo {author} {\bibfnamefont {V.}~\bibnamefont {Thakore}},
  \bibinfo {author} {\bibfnamefont {F.}~\bibnamefont {Seyedheydari}}, \bibinfo
  {author} {\bibfnamefont {M.}~\bibnamefont {Karttunen}}, \ and\ \bibinfo
  {author} {\bibfnamefont {T.}~\bibnamefont {Ala-Nissila}},\ }\href@noop {}
  {\bibfield  {journal} {\bibinfo  {journal} {AIP Advances}\ }\textbf {\bibinfo
  {volume} {10}},\ \bibinfo {pages} {095128} (\bibinfo {year}
  {2020})}\BibitemShut {NoStop}%
\bibitem [{\citenamefont {Tang}\ \emph {et~al.}(2017)\citenamefont {Tang},
  \citenamefont {Thakore},\ and\ \citenamefont
  {Ala-Nissila}}]{tang2017plasmonically}%
  \BibitemOpen
  \bibfield  {author} {\bibinfo {author} {\bibfnamefont {J.}~\bibnamefont
  {Tang}}, \bibinfo {author} {\bibfnamefont {V.}~\bibnamefont {Thakore}}, \
  and\ \bibinfo {author} {\bibfnamefont {T.}~\bibnamefont {Ala-Nissila}},\
  }\href@noop {} {\bibfield  {journal} {\bibinfo  {journal} {Sci Rep.}\
  }\textbf {\bibinfo {volume} {7}},\ \bibinfo {pages} {5696} (\bibinfo {year}
  {2017})}\BibitemShut {NoStop}%
\bibitem [{\citenamefont {Slovick}\ \emph {et~al.}(2015)\citenamefont
  {Slovick}, \citenamefont {Baker}, \citenamefont {Flom},\ and\ \citenamefont
  {Krishnamurthy}}]{slovick2015tailoring}%
  \BibitemOpen
  \bibfield  {author} {\bibinfo {author} {\bibfnamefont {B.~A.}\ \bibnamefont
  {Slovick}}, \bibinfo {author} {\bibfnamefont {J.~M.}\ \bibnamefont {Baker}},
  \bibinfo {author} {\bibfnamefont {Z.}~\bibnamefont {Flom}}, \ and\ \bibinfo
  {author} {\bibfnamefont {S.}~\bibnamefont {Krishnamurthy}},\ }\href@noop {}
  {\bibfield  {journal} {\bibinfo  {journal} {Appl. Phys. Lett.}\ }\textbf
  {\bibinfo {volume} {107}},\ \bibinfo {pages} {141903} (\bibinfo {year}
  {2015})}\BibitemShut {NoStop}%
\bibitem [{\citenamefont {Kerker}(2013)}]{kerker2013scattering}%
  \BibitemOpen
  \bibfield  {author} {\bibinfo {author} {\bibfnamefont {M.}~\bibnamefont
  {Kerker}},\ }\enquote {\bibinfo {title} {Scattering by a sphere},}\ in\
  \href@noop {} {\emph {\bibinfo {booktitle} {The scattering of light and other
  electromagnetic radiation: {P}hysical chemistry: {A} series of
  monographs}}},\ Vol.~\bibinfo {volume} {16}\ (\bibinfo  {publisher} {Academic
  press},\ \bibinfo {year} {2013})\ pp.\ \bibinfo {pages} {27--98}\BibitemShut
  {NoStop}%
\bibitem [{\citenamefont {Alaee}\ \emph {et~al.}(2015)\citenamefont {Alaee},
  \citenamefont {Filter}, \citenamefont {Lehr}, \citenamefont {Lederer},\ and\
  \citenamefont {Rockstuhl}}]{alaee2015generalized}%
  \BibitemOpen
  \bibfield  {author} {\bibinfo {author} {\bibfnamefont {R.}~\bibnamefont
  {Alaee}}, \bibinfo {author} {\bibfnamefont {R.}~\bibnamefont {Filter}},
  \bibinfo {author} {\bibfnamefont {D.}~\bibnamefont {Lehr}}, \bibinfo {author}
  {\bibfnamefont {F.}~\bibnamefont {Lederer}}, \ and\ \bibinfo {author}
  {\bibfnamefont {C.}~\bibnamefont {Rockstuhl}},\ }\href@noop {} {\bibfield
  {journal} {\bibinfo  {journal} {Opt. Lett.}\ }\textbf {\bibinfo {volume}
  {40}},\ \bibinfo {pages} {2645} (\bibinfo {year} {2015})}\BibitemShut
  {NoStop}%
\bibitem [{\citenamefont {Kruk}\ and\ \citenamefont
  {Kivshar}(2017)}]{kruk2017functional}%
  \BibitemOpen
  \bibfield  {author} {\bibinfo {author} {\bibfnamefont {S.}~\bibnamefont
  {Kruk}}\ and\ \bibinfo {author} {\bibfnamefont {Y.}~\bibnamefont {Kivshar}},\
  }\href@noop {} {\bibfield  {journal} {\bibinfo  {journal} {ACS Photonics}\
  }\textbf {\bibinfo {volume} {4}},\ \bibinfo {pages} {2638} (\bibinfo {year}
  {2017})}\BibitemShut {NoStop}%
\bibitem [{\citenamefont {Maduraiveeran}\ and\ \citenamefont
  {Ramaraj}(2007)}]{maduraiveeran2007gold}%
  \BibitemOpen
  \bibfield  {author} {\bibinfo {author} {\bibfnamefont {G.}~\bibnamefont
  {Maduraiveeran}}\ and\ \bibinfo {author} {\bibfnamefont {R.}~\bibnamefont
  {Ramaraj}},\ }\href@noop {} {\bibfield  {journal} {\bibinfo  {journal} {J.
  Electroanal. Chem.}\ }\textbf {\bibinfo {volume} {608}},\ \bibinfo {pages}
  {52} (\bibinfo {year} {2007})}\BibitemShut {NoStop}%
\bibitem [{\citenamefont {Cuevas}\ \emph {et~al.}(2000)\citenamefont {Cuevas},
  \citenamefont {Gusken}, \citenamefont {Sekiya}, \citenamefont {Ogata},
  \citenamefont {Torikai},\ and\ \citenamefont {Suzuki}}]{cuevas2000effect}%
  \BibitemOpen
  \bibfield  {author} {\bibinfo {author} {\bibfnamefont {R.~F.}\ \bibnamefont
  {Cuevas}}, \bibinfo {author} {\bibfnamefont {E.}~\bibnamefont {Gusken}},
  \bibinfo {author} {\bibfnamefont {E.~H.}\ \bibnamefont {Sekiya}}, \bibinfo
  {author} {\bibfnamefont {D.~Y.}\ \bibnamefont {Ogata}}, \bibinfo {author}
  {\bibfnamefont {D.}~\bibnamefont {Torikai}}, \ and\ \bibinfo {author}
  {\bibfnamefont {C.~K.}\ \bibnamefont {Suzuki}},\ }\href@noop {} {\bibfield
  {journal} {\bibinfo  {journal} {J. Non-Cryst. Solids}\ }\textbf {\bibinfo
  {volume} {273}},\ \bibinfo {pages} {252} (\bibinfo {year}
  {2000})}\BibitemShut {NoStop}%
\bibitem [{\citenamefont {Hiratsuka}\ \emph {et~al.}(2007)\citenamefont
  {Hiratsuka}, \citenamefont {Tatami}, \citenamefont {Wakihara}, \citenamefont
  {Komeya},\ and\ \citenamefont {Meguro}}]{hiratsuka2007fabrication}%
  \BibitemOpen
  \bibfield  {author} {\bibinfo {author} {\bibfnamefont {D.}~\bibnamefont
  {Hiratsuka}}, \bibinfo {author} {\bibfnamefont {J.}~\bibnamefont {Tatami}},
  \bibinfo {author} {\bibfnamefont {T.}~\bibnamefont {Wakihara}}, \bibinfo
  {author} {\bibfnamefont {K.}~\bibnamefont {Komeya}}, \ and\ \bibinfo {author}
  {\bibfnamefont {T.}~\bibnamefont {Meguro}},\ }\href@noop {} {\bibfield
  {journal} {\bibinfo  {journal} {J. Ceram. Soc. Jpn.}\ }\textbf {\bibinfo
  {volume} {115}},\ \bibinfo {pages} {392} (\bibinfo {year}
  {2007})}\BibitemShut {NoStop}%
\bibitem [{\citenamefont {Szafran}\ \emph {et~al.}(2007)\citenamefont
  {Szafran}, \citenamefont {Konopka}, \citenamefont {Bobryk},\ and\
  \citenamefont {Kurzyd{\l}owski}}]{szafran2007ceramic}%
  \BibitemOpen
  \bibfield  {author} {\bibinfo {author} {\bibfnamefont {M.}~\bibnamefont
  {Szafran}}, \bibinfo {author} {\bibfnamefont {K.}~\bibnamefont {Konopka}},
  \bibinfo {author} {\bibfnamefont {E.}~\bibnamefont {Bobryk}}, \ and\ \bibinfo
  {author} {\bibfnamefont {K.}~\bibnamefont {Kurzyd{\l}owski}},\ }\href@noop {}
  {\bibfield  {journal} {\bibinfo  {journal} {J. Eur. Ceram. Soc.}\ }\textbf
  {\bibinfo {volume} {27}},\ \bibinfo {pages} {651} (\bibinfo {year}
  {2007})}\BibitemShut {NoStop}%
\bibitem [{\citenamefont {Munir}\ \emph {et~al.}(2011)\citenamefont {Munir},
  \citenamefont {Quach},\ and\ \citenamefont {Ohyanagi}}]{munir2011electric}%
  \BibitemOpen
  \bibfield  {author} {\bibinfo {author} {\bibfnamefont {Z.~A.}\ \bibnamefont
  {Munir}}, \bibinfo {author} {\bibfnamefont {D.~V.}\ \bibnamefont {Quach}}, \
  and\ \bibinfo {author} {\bibfnamefont {M.}~\bibnamefont {Ohyanagi}},\
  }\href@noop {} {\bibfield  {journal} {\bibinfo  {journal} {J. Am. Ceram.
  Soc.}\ }\textbf {\bibinfo {volume} {94}},\ \bibinfo {pages} {1} (\bibinfo
  {year} {2011})}\BibitemShut {NoStop}%
\bibitem [{\citenamefont {Singh}\ \emph {et~al.}(2014)\citenamefont {Singh},
  \citenamefont {Cura}, \citenamefont {Liu}, \citenamefont {Johansson},
  \citenamefont {Ge},\ and\ \citenamefont {Hannula}}]{singh2014tuning}%
  \BibitemOpen
  \bibfield  {author} {\bibinfo {author} {\bibfnamefont {V.~K.}\ \bibnamefont
  {Singh}}, \bibinfo {author} {\bibfnamefont {M.~E.}\ \bibnamefont {Cura}},
  \bibinfo {author} {\bibfnamefont {X.}~\bibnamefont {Liu}}, \bibinfo {author}
  {\bibfnamefont {L.-S.}\ \bibnamefont {Johansson}}, \bibinfo {author}
  {\bibfnamefont {Y.}~\bibnamefont {Ge}}, \ and\ \bibinfo {author}
  {\bibfnamefont {S.-P.}\ \bibnamefont {Hannula}},\ }\href@noop {} {\bibfield
  {journal} {\bibinfo  {journal} {ChemPlusChem}\ }\textbf {\bibinfo {volume}
  {79}},\ \bibinfo {pages} {1512} (\bibinfo {year} {2014})}\BibitemShut
  {NoStop}%
\bibitem [{\citenamefont {He}\ \emph {et~al.}(2013)\citenamefont {He},
  \citenamefont {Tu}, \citenamefont {Katsui},\ and\ \citenamefont
  {Goto}}]{he2013synthesis}%
  \BibitemOpen
  \bibfield  {author} {\bibinfo {author} {\bibfnamefont {Z.}~\bibnamefont
  {He}}, \bibinfo {author} {\bibfnamefont {R.}~\bibnamefont {Tu}}, \bibinfo
  {author} {\bibfnamefont {H.}~\bibnamefont {Katsui}}, \ and\ \bibinfo {author}
  {\bibfnamefont {T.}~\bibnamefont {Goto}},\ }\href@noop {} {\bibfield
  {journal} {\bibinfo  {journal} {Ceram. Int.}\ }\textbf {\bibinfo {volume}
  {39}},\ \bibinfo {pages} {2605} (\bibinfo {year} {2013})}\BibitemShut
  {NoStop}%
\bibitem [{SI()}]{SI}%
  \BibitemOpen
  \href@noop {} {\enquote {\bibinfo {title} {See {S}upplemental {M}aterial at
  [{URL} to be inserted by publisher] for experimental characterization and
  details of the computational methods.}}\ }\BibitemShut {NoStop}%
\bibitem [{\citenamefont {Wang}\ \emph {et~al.}(1995)\citenamefont {Wang},
  \citenamefont {Jacques},\ and\ \citenamefont {Zheng}}]{wang1995mcml}%
  \BibitemOpen
  \bibfield  {author} {\bibinfo {author} {\bibfnamefont {L.}~\bibnamefont
  {Wang}}, \bibinfo {author} {\bibfnamefont {S.~L.}\ \bibnamefont {Jacques}}, \
  and\ \bibinfo {author} {\bibfnamefont {L.}~\bibnamefont {Zheng}},\
  }\href@noop {} {\bibfield  {journal} {\bibinfo  {journal} {Comput Methods
  Programs Biomed.}\ }\textbf {\bibinfo {volume} {47}},\ \bibinfo {pages} {131}
  (\bibinfo {year} {1995})}\BibitemShut {NoStop}%
\bibitem [{\citenamefont {Palik}(1998)}]{palik1998handbook}%
  \BibitemOpen
  \bibfield  {author} {\bibinfo {author} {\bibfnamefont {E.~D.}\ \bibnamefont
  {Palik}},\ }\href@noop {} {\emph {\bibinfo {title} {Handbook of optical
  constants of solids}}},\ Vol.~\bibinfo {volume} {3}\ (\bibinfo  {publisher}
  {Academic press},\ \bibinfo {year} {1998})\BibitemShut {NoStop}%
\bibitem [{\citenamefont {Merikallio}(2016)}]{merikallio2016computer}%
  \BibitemOpen
  \bibfield  {author} {\bibinfo {author} {\bibfnamefont {S.}~\bibnamefont
  {Merikallio}},\ }\emph {\bibinfo {title} {Computer modeling of light
  scattering by atmospheric dust particles with spheroids and ellipsoids}},\
  \href@noop {} {Ph.D. thesis},\ \bibinfo  {school} {Aalto University}
  (\bibinfo {year} {2016})\BibitemShut {NoStop}%
\bibitem [{\citenamefont {Markel}(2016)}]{markel2016introduction}%
  \BibitemOpen
  \bibfield  {author} {\bibinfo {author} {\bibfnamefont {V.~A.}\ \bibnamefont
  {Markel}},\ }\href@noop {} {\bibfield  {journal} {\bibinfo  {journal} {J.
  Opt. Soc. Am. A}\ }\textbf {\bibinfo {volume} {33}},\ \bibinfo {pages} {1244}
  (\bibinfo {year} {2016})}\BibitemShut {NoStop}%
\bibitem [{\citenamefont {Kubelka}(1948)}]{kubelka1948new}%
  \BibitemOpen
  \bibfield  {author} {\bibinfo {author} {\bibfnamefont {P.}~\bibnamefont
  {Kubelka}},\ }\href@noop {} {\bibfield  {journal} {\bibinfo  {journal} {J.
  Opt. Soc. Am.}\ }\textbf {\bibinfo {volume} {38}},\ \bibinfo {pages} {448}
  (\bibinfo {year} {1948})}\BibitemShut {NoStop}%
\bibitem [{\citenamefont {{v}an Gemert}\ \emph {et~al.}(1987)\citenamefont
  {{v}an Gemert}, \citenamefont {Welch}, \citenamefont {Star}, \citenamefont
  {Motamedi},\ and\ \citenamefont {Cheong}}]{van1987tissue}%
  \BibitemOpen
  \bibfield  {author} {\bibinfo {author} {\bibfnamefont {M.~J.}\ \bibnamefont
  {{v}an Gemert}}, \bibinfo {author} {\bibfnamefont {A.~J.}\ \bibnamefont
  {Welch}}, \bibinfo {author} {\bibfnamefont {W.~M.}\ \bibnamefont {Star}},
  \bibinfo {author} {\bibfnamefont {M.}~\bibnamefont {Motamedi}}, \ and\
  \bibinfo {author} {\bibfnamefont {W.-F.}\ \bibnamefont {Cheong}},\
  }\href@noop {} {\bibfield  {journal} {\bibinfo  {journal} {Laser. Med. Sci.}\
  }\textbf {\bibinfo {volume} {2}},\ \bibinfo {pages} {295} (\bibinfo {year}
  {1987})}\BibitemShut {NoStop}%
\bibitem [{\citenamefont {Murphy}(2006)}]{murphy2006modified}%
  \BibitemOpen
  \bibfield  {author} {\bibinfo {author} {\bibfnamefont {A.}~\bibnamefont
  {Murphy}},\ }\href@noop {} {\bibfield  {journal} {\bibinfo  {journal} {J.
  Phys. D}\ }\textbf {\bibinfo {volume} {39}},\ \bibinfo {pages} {3571}
  (\bibinfo {year} {2006})}\BibitemShut {NoStop}%
\bibitem [{\citenamefont {Tzarouchis}\ \emph {et~al.}(2016)\citenamefont
  {Tzarouchis}, \citenamefont {Yl{\"a}-Oijala}, \citenamefont {Ala-Nissila},\
  and\ \citenamefont {Sihvola}}]{tzarouchis2016shape}%
  \BibitemOpen
  \bibfield  {author} {\bibinfo {author} {\bibfnamefont {D.~C.}\ \bibnamefont
  {Tzarouchis}}, \bibinfo {author} {\bibfnamefont {P.}~\bibnamefont
  {Yl{\"a}-Oijala}}, \bibinfo {author} {\bibfnamefont {T.}~\bibnamefont
  {Ala-Nissila}}, \ and\ \bibinfo {author} {\bibfnamefont {A.}~\bibnamefont
  {Sihvola}},\ }\href@noop {} {\bibfield  {journal} {\bibinfo  {journal}
  {Applied Physics A}\ }\textbf {\bibinfo {volume} {122}},\ \bibinfo {pages}
  {298} (\bibinfo {year} {2016})}\BibitemShut {NoStop}%
\bibitem [{\citenamefont {Vargas}\ and\ \citenamefont
  {Niklasson}(1997)}]{vargas1997applicability}%
  \BibitemOpen
  \bibfield  {author} {\bibinfo {author} {\bibfnamefont {W.~E.}\ \bibnamefont
  {Vargas}}\ and\ \bibinfo {author} {\bibfnamefont {G.~A.}\ \bibnamefont
  {Niklasson}},\ }\href@noop {} {\bibfield  {journal} {\bibinfo  {journal}
  {Appl. Opt.}\ }\textbf {\bibinfo {volume} {36}},\ \bibinfo {pages} {5580}
  (\bibinfo {year} {1997})}\BibitemShut {NoStop}%
\bibitem [{\citenamefont {Conley}\ \emph {et~al.}(2018)\citenamefont {Conley},
  \citenamefont {Thakore},\ and\ \citenamefont
  {Ala-Nissila}}]{conley2018plasmonically}%
  \BibitemOpen
  \bibfield  {author} {\bibinfo {author} {\bibfnamefont {K.}~\bibnamefont
  {Conley}}, \bibinfo {author} {\bibfnamefont {V.}~\bibnamefont {Thakore}}, \
  and\ \bibinfo {author} {\bibfnamefont {T.}~\bibnamefont {Ala-Nissila}},\ }in\
  \href@noop {} {\emph {\bibinfo {booktitle} {2018 Progress in Electromagnetics
  Research Symposium (PIERS-Toyama)}}}\ (\bibinfo {organization} {IEEE},\
  \bibinfo {year} {2018})\ pp.\ \bibinfo {pages} {2435--2441}\BibitemShut
  {NoStop}%
\end{thebibliography}%

\end{document}